\documentclass[]{spie}  

\usepackage{amsmath,amsfonts,amssymb}
\usepackage{graphicx}
\usepackage[colorlinks=true, allcolors=blue]{hyperref}
\usepackage{cite}

\title{CCAT: Nonlinear effects in 280 GHz aluminum kinetic inductance detectors}

\author[a]{Cody J. Duell}
\author[b]{Jason Austermann}
\author[c]{James R. Burgoyne}
\author[c,d,e]{Scott C. Chapman}
\author[f]{Steve K. Choi}
\author[a]{Abigail T. Crites}
\author[g]{Rodrigo G. Freundt}
\author[h]{Anthony I. Huber}
\author[a]{Zachary B. Huber}
\author[b]{Johannes Hubmayr}
\author[a]{Ben Keller}
\author[a]{Lawrence T. Lin}
\author[a]{Alicia M. Middleton}
\author[a]{Colin C. Murphy}
\author[a,g]{Michael D. Niemack}
\author[i]{Thomas Nikola}
\author[a]{Darshan Patel}
\author[c]{Adrian K. Sinclair}
\author[a]{Ema Smith}
\author[g]{Gordon J. Stacey}
\author[b]{Anna Vaskuri}
\author[j,a]{Eve M. Vavagiakis}
\author[b]{Michael Vissers}
\author[a]{Samantha Walker}
\author[b]{Jordan Wheeler}

\affil[a]{Department of Physics, Cornell University, Ithaca, NY, 14853, USA}
\affil[b]{Quantum Sensors Division, National Institute of Standards and Technology, Boulder, CO, 80305, USA}
\affil[c]{Department of Physics and Astronomy, University of British Columbia, Vancouver, BC, V6T 1Z4, Canada }
\affil[d]{Department of Physics and Atmospheric Science, Dalhousie University, Halifax, NS, B3H 4R2, Canada}
\affil[e]{National Research Council, Herzberg Astronomy and Astrophysics, Victoria, BC, V9E 2E7, Canada}
\affil[f]{Department of Physics and Astronomy, University of California, Riverside, CA, 92521, USA}
\affil[g]{Department of Astronomy, Cornell University, Ithaca, NY, 14853, USA}
\affil[h]{Department of Physics and Astronomy, University of Victoria, Victoria, BC, V8P 5C2, Canada }
\affil[i]{Cornell Center for Astrophysics and Planetary Sciences, Cornell University, Ithaca, NY, 14853, USA}
\affil[j]{Department of Physics, Duke University, Durham, NC, 27708, USA}

\authorinfo{Send correspondence to C. Duell, E-mail: cjd259@cornell.edu}

\begin{document} 
\maketitle
\begin{abstract}
Prime-Cam, a first-generation science instrument for the Atacama-based Fred Young Submillimeter Telescope, is being built by the CCAT Collaboration to observe at millimeter and submillimeter wavelengths using kinetic inductance detectors (KIDs). Prime-Cam’s 280 GHz instrument module will deploy with two aluminum-based KID arrays and one titanium nitride-based KID array, totaling $\sim$10,000 detectors at the focal plane, all of which have been fabricated and are currently undergoing testing. One complication of fielding large arrays of KIDs under dynamic loading conditions is tuning the detector tone powers to maximize signal-to-noise while avoiding bifurcation due to the nonlinear kinetic inductance. For aluminum-based KIDs, this is further complicated by additional nonlinear effects which couple tone power to resonator quality factors and resonant frequencies. While both nonequilibrium quasiparticle dynamics and two-level system fluctuations have been shown to give rise to qualitatively similar distortions, modeling these effects alongside nonlinear kinetic inductance is inefficient when fitting thousands of resonators on-sky with existing models. For this reason, it is necessary to have a detailed understanding of the nonlinear effects across relevant detector loading conditions, including how they impact on on-sky noise and how to diagnose the detector’s relative performance. We present a study of the competing nonlinearities seen in Prime-Cam’s 280 GHz aluminum KIDs, with a particular emphasis on the resulting distortions to the resonator line shape and how these impact detector parameter estimation.   
\end{abstract}


\section{Introduction}
\label{sec:intro}

Kinetic inductance detectors (KIDs), a type of frequency-division-multiplexed superconducting resonators \cite{day_broadband_2003}, have become an increasingly popular choice in recent years for observing at millimeter and submillimeter wavelengths \cite{monfardini_dual-band_2011, adam_nika2_2018, paiella_kinetic_2019, brien_muscat_2018, dober_next-generation_2014, wilson_toltec_2020}. The CCAT collaboration's Prime-Cam instrument \cite{vavagiakis_prime-cam_2018, choi_sensitivity_2020} will use KIDs for observing at the Atacama-based Fred Young Submillimeter Telescope (FYST), beginning with three arrays (one TiN and two Al) at 280 GHz. This first module will be followed in the near-term by imaging modules at 350 GHz and 850 GHz, and a spectrometer module, EoR-Spec\cite{collaboration_ccat-prime_2022, choi_sensitivity_2020}. As we approach deployment, one of the practical challenges for operation is the distinctive nonlinear response of aluminum KIDs. Resonator nonlinearity occurs when the underlying circuit parameters are sensitive to the internal energy of the resonator. When nearing resonance under these conditions, the increasing absorption of microwave probe tone photons leads to a changing resonator profile even at fixed tone powers. 

While many strategies for tone power optimization are built around the well-understood nonlinear kinetic inductance of TiN\cite{swenson_operation_2013}, aluminum KIDs can exhibit additional competing nonlinearity as a result of nonequilibrium quasiparticle dynamics \cite{de_visser_non-equilibrium_2015}. In Prime-Cam's 280 GHz aluminum KIDs, these competing nonlinearities distort the resonator line shapes even well-below the onset of bifurcation, skewing resonator fits and increasing the noise penalty for sub-optimal tone powers or tone placement. By using information from the full resonator sweep, we can observe the evolution of this detector nonlinearity and better understand the impact on resonator fits and tone optimization.

\section{Background}\label{sec:methods}

As previously described\cite{choi_ccat-prime_2022, austermann_aluminum-based_2022, duell_ccat_2024}, the first KID arrays for Prime-Cam have already completed production and begun in-lab characterization. The data described here was taken using three witness pixels (five total detectors) fabricated on the same wafer as the first completed Al array in a set-up that allows for measurement with a cryogenic black-body source at a range of base temperatures down to $\sim 58$~mK. An extensive amount of data has been acquired with the Al witness pixels under a variety of bath temperature, optical loading, and probe tone power conditions. 

\begin{figure}[htbp]%
\centering
\includegraphics[width=0.7\textwidth]{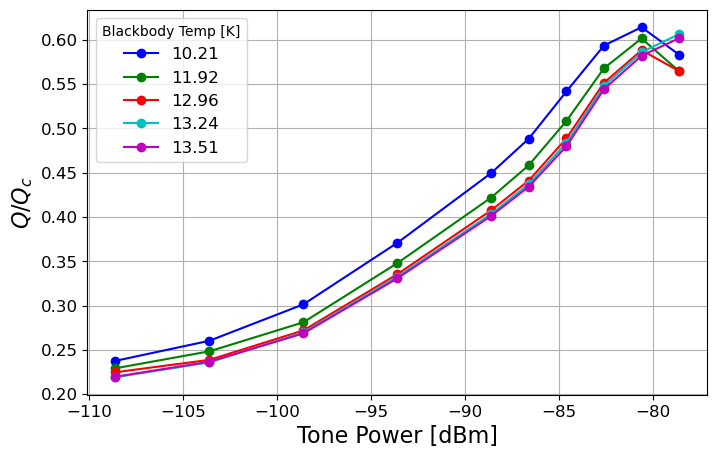}
\caption[The ratio of the $Q/Q_c$ as a function of raw probe tone power in dBm.]{The ratio of the $Q/Q_c$ as a function of estimated probe tone power in dBm. Different colors correspond to cryogenic black-body temperatures and fits at the highest tone powers are somewhat unreliable. The noteworthy aspect here for tone-power optimization is that at each load temperature the changing tone power drives the resonator from being $Q_i$-dominated ($Q/Q_c<0.5$) to $Q_c$-dominated ($Q/Q_c>0.5$). Resonator noise is higher when operating in the $Q_i$-dominated regime.}\label{fig:al_q_rat}
\end{figure}

All five measured resonators are between 500 MHz and 901 MHz. Measured coupling quality factors ($Q_c$) are in the range of 18,000 to 36,000, and under designed loading conditions, the total quality factor ($Q$) values are expected to be in the range of 8,000 to 20,000. As is discussed further in section \ref{sec5}, tone power optimization has a significant impact on both the measurable resonator parameters and the observed shape of the resonance circle, particularly at tone powers that are most relevant for operation. Since these effects on the line shape are not easily modeled, they significantly skew any resonator fits to systematically underestimate quality factors and rotate the impedance mismatch angle. Figure \ref{fig:al_q_rat} shows how going from low powers to higher powers pushes the KID from deeply in the internal quality factor ($Q_i$) dominated regime ($Q/Q_c < 0.5$) through critical coupling ($Q/Q_c = 0.5$) to the $Q_c$-dominated regime  ($Q/Q_c > 0.5$).

\begin{figure}[htbp]
   \begin{center}
   \includegraphics[width=\textwidth]{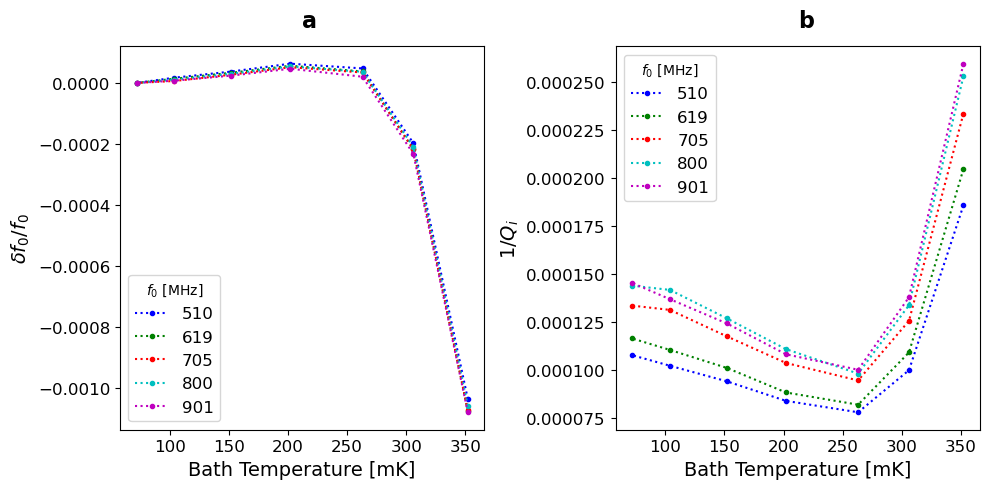}
   \end{center}
\caption{(a) Fractional frequency shift vs. bath temperature based on fitted data from five Al detectors under a 3~K blackbody load and $\sim15$ dB below bifurcation for ease of fitting. (b) Internal loss ($Q_i^{-1}$) vs. bath temperature for the same five resonators. While both parameters are sensitive to tone power, the quality factors are much more-so, causing the relatively wide spread between these detectors.}
\label{fig:Al_fit_data} 
\end{figure}

In addition to the tone-power sensitivity, the Al detectors deviate from the equilibrium Mattis-Bardeen evolution with bath temperature by first increasing in resonance frequency ($f_0$) and $Q_i$ before turning around to a more standard continuous decrease in both parameters. Figure \ref{fig:Al_fit_data} shows this resonator response to changing bath temperatures based on well-fit data, which is, by necessity, $15$-$20$ dB below the optimal operating powers of the resonators. This "under-driving" of the resonators shows most significantly in the plots of the internal resonator loss, $Q_i^{-1}$ (also sometimes written as tan$\delta_i$), which is much larger and much less consistent between KIDs than it would be under optimal tone powers.

\section{Detector Nonlinearity}\label{sec5}

When describing resonator behavior, either linear or nonlinear, we are describing a single-pole Lorentzian that is parameterized by the center resonant frequency ($f_0$) and the total quality factor ($Q$). For the particular case of a capacitively-coupled resonator, the forward scattering parameter, $S_{21}$, takes the form 
\begin{equation}
S_{21} = 1 - \frac{Q}{Q_c} \frac{1}{1 + 2jQ\frac{f-f_0}{f_0}}= 1 - \frac{Q}{Q_c} \frac{1}{1 + 2jQx} \text{,}
\end{equation}
where $Q_c$ is the coupling quality factor, $f$ is the probe frequency, and $x \equiv \frac{f-f_0}{f_0}$. In the complex, or Argand, plane this traces out a circle of diameter $Q/Q_c$ that is centered at $(1 - \frac{Q}{2 Q_c}, 0)$. Given the definitions of $Q$ and $S_{21}$, we can write the resonator's internal energy, $E_r$, as
\begin{equation}
    E_r = \frac{2Q^2}{Q_c}\frac{1}{1+4Q^2 x^2}\frac{P_r}{2\pi f_0} \text{,}
\end{equation}
where $P_r$ is the readout tone power. Critically, the probe tone appears here through both the tone power ($P_r$) and the frequency (through $x=\frac{f-f_0}{f_0}$). In the case of a linear resonator, the parameters $f_0$ and $Q$ are stationary and each point on the resonance circle can be mapped back to the same values. If instead there is some dependance of circuit parameters on the internal energy, then we can see non-linear behavior where the underlying parameters are changing along with the tone power or frequency. This can be driven by a number of different underlying physical processes (see \ref{subsec:NL_types}) and can give rise to a wide range of behaviors, many of which are described in \citenum{thomas_nonlinear_2020}. In general, however, we can refer to nonlinearities as either reactive (where $f_0$ is changing), dissipative (where $Q$ is changing), or both. Reactive nonlinearities shift the position (often just referred to as the phase) on the resonance circle, while dissipative nonlinearities change the diameter of the circle and the circle-phase relationship to $f$.

\subsection{Types of Nonlinearities}\label{subsec:NL_types}

The most well-understood form of nonlinearity in KIDs is the nonlinear kinetic inductance\cite{swenson_operation_2013}, which results from a higher-order dependence on current in the kinetic inductance. This is a purely reactive nonlinearity, and, when viewed on the resonance circle at higher powers, appears as a lopsided jump in phase with no change in the circle diameter. The asymmetry is caused by $f_0$ shifting monotonically to lower frequencies as the resonator energy increases. As $f$ approaches $f_0$ from below ($x<0$), $x^2$ decreases and $E_r$ rises, pulling $f_0$ down in frequency at an increasing rate until eventually we jump over the center frequency and $x^2$ begins to increase again. On the high frequency side of the resonance, where $x>0$, the resonator relaxes back into its higher frequency state as $E_r$ decreases, following the probe tone and causing the phase to evolve more slowly than it would otherwise. Most critically, at high tone powers the nonlinear kinetic inductance causes the resonator state to "bifurcate" as the system discontinuously jumps between different possible resonator states, setting a practical upper limit on tone powers.

Another source of nonlinearity with a well-defined impact is loss to two-level systems (TLS) \cite{gao_semiempirical_2008, pappas_two_2011} in the dielectric. At very low temperatures and tone powers, TLS fluctuations increase the overall loss, decreasing the $Q$. Crucially, this effect is reduced at higher temperatures and powers. The latter of these allows for nonlinear feedback. As the power stored in the TLS increases, the $Q$ likewise increases such that the circle diameter reaches a maximum on resonance. This can cause the resonance circle to take on an oblong shape, appearing squashed on the sides. This effect is generally quite small by design, as the total loss is the sum of several contributions, chiefly the coupling loss (loss to the readout line), the loss to the quasiparticle system, and the TLS loss. Detectors are designed to operate in regimes dominated by coupling loss or quasiparticle loss. 

Finally, quasiparticle absorption of microwave photons has been shown to lead to strongly nonlinear behavior \cite{de_visser_evidence_2014, de_visser_non-equilibrium_2015}. By driving quasiparticles out of thermal equilibrium, sub-gap microwave photons push the system away from the expected Fermi-Dirac distribution, as well as explicitly altering the density of states. Both of these appear when calculating the AC conductivity from the Mattis-Bardeen equations, allowing for tone power to alter the conductivity in both reactive and dissipative manners. While calculating these effects is beyond the scope of this work, we briefly describe the two scenarios that can be observed, which correspond to an effective "heating" or "cooling" of the quasiparticle system. For more detailed discussion of this approach, refer to \citenum{goldie_non-equilibrium_2012, guruswamy_quasiparticle_2014, thomas_electrothermal_2015, semenov_coherent_2016}. It should be noted that, while these scenarios are observed in \citenum{de_visser_evidence_2014} to occur at different temperature regimes, they  are sufficiently complicated that either "heating" or "cooling" can occur at a wide range of temperatures.  In the case of quasiparticle cooling, quasiparticles are excited by the probe tone to recombine more rapidly than they would otherwise, causing a reduction in both loss (increase in $Q$) and kinetic inductance (increase in $f_0$). The increase in $Q$ causes the diameter of the circle to increase as you approach resonance, while the increase in $f_0$ causes the probe frequency to approach resonance more slowly from below and more rapidly from above, leading to an asymmetry. In the case of quasiparticle heating, the excess energy in the quasiparticle system instead suppresses recombination, causing an increase in both loss and inductance in a manner similar to above-gap pair-breaking radiation. Decreasing the $Q$ in this case squashes the resonance circle in the opposite manner to above, meaning that the diameter is at a minimum when on resonance. Decreasing the resonance frequency causes an effect similar to the nonlinear kinetic inductance, resulting in an asymmetry in the resonance circle of the same manner. 

Since the kinetic inductance nonlinearity is monotonic and well-understood, we can account for it in a relatively straightforward manner when fitting a resonator\cite{swenson_operation_2013, walker_development_2023}. This is the case for Prime-Cam's TiN KIDs\cite{choi_ccat-prime_2022}. In cases where multiple observable nonlinearities arise from  different, potentially opposing physical effects, this modeling becomes more difficult. For the Al KIDs these effects are severe enough to significantly alter the observed resonator profile and bias fits using either a standard linear model or a model incorporating nonlinear kinetic inductance. As such, it is useful to "unwrap" the resonance circle to show what nonlinearities most significantly impact the observed resonator shape. 

\section{Unwrapping the Resonance Circle}

When we take an $S_{21}$ trace near resonance, the equation that we are effectively measuring as a function of frequency $\omega = \omega_0 (1 + x)$ is \cite{mauskopf_transition_2018}
\begin{equation}\label{eq:full_s21}
    S_{21} = A(\omega) e^{j\theta(\omega)} \left(1 - \frac{Q}{Q_e}\frac{1}{1 + 2 j Q x}\right)\text{,}
\end{equation}
where $A(\omega)e^{j\theta(\omega)}$ is a frequency-dependent complex normalization and $Q_e$ is a generalized complex version of $Q_c$ that allows for an additional rotation due to mismatches in the resonator and feedline impedance.
Beginning from a measurement of $S_{21}$, if we remove the factors that arise from the environment (i.e. the cables, amplifiers, etc.) and the impedance mismatch (the rotation angle of the external quality factor, $Q_e$), we can arrive back at the pure resonator expression. Shifting our anchor point to the origin for convenience, we can rewrite this expression (which we call $\hat{s}_{res}$) in terms of the two parameters that ought to be linear:
\begin{equation}\label{eq:s_hat}
\begin{aligned}
    \hat{s}_{res} =  1 - S_{21, res} &= \frac{Q}{|Q_e|} \frac{1}{1 + 2jQ\frac{\omega-\omega_0}{\omega_0}}\\
    &= \frac{q_r(\omega)}{1+2jy(\omega)} \text{ .}
\end{aligned}
\end{equation}
Here, $q_r = Q/|Q_e|$ is the diameter of the resonance circle (related to the dip depth) and $y = Q\frac{\omega-\omega_0}{\omega_0} = Q x$ is the distance from the center frequency as measured in line-widths. For a linear resonator, $q_r$ will be constant and $y$ will be linear in frequency with the slope set by the $Q$ and the zero value set by $\omega_0 = 2\pi f_0$. With the environment and impedance mismatch accounted for, we can convert a position in the complex plane to an implied $q_r$ and $y$, thus, we can use these to fit for $Q$, $f_0$, and $|Q_e|$. From Equation \ref{eq:s_hat}, we can calculate
\begin{equation}\label{eq:yf}
y(\omega) = -\frac{1}{2}\frac{\mathfrak{Im}(\hat{s}_{res})}{\mathfrak{Re}(\hat{s}_{res})}
\end{equation}
and
\begin{equation}\label{eq:Af}
q_r(\omega) = \left[1 + \left(\frac{\mathfrak{Im}(\hat{s}_{res})}{\mathfrak{Re}(\hat{s}_{res})}\right)^2\right]\mathfrak{Re}(\hat{s}_{res})\text{,}
\end{equation}
where $\mathfrak{Re}(\hat{s}_{res})$ and $\mathfrak{Im}(\hat{s}_{res})$ are the real and imaginary parts of $\hat{s}_{res}$, respectively. Using these expressions to identify trends in $Q$ and $f_0$ does require a good estimation of and proper accounting for environmental effects. Additionally, the appearance of  $\mathfrak{Re}(\hat{s}_{res})$ in the denominator of both Equation \ref{eq:yf} and Equation \ref{eq:Af} means that the impacts of noise on calculated parameters becomes much larger for smaller resonance circles or farther away from resonance. 

One final improvement to separating out reactive and dissipative nonlinearities in these plots is to plot $y_c = y/q_r = Q_c\frac{\omega-\omega_0}{\omega_0}$ rather than $y$, since $Q_c$ generally should not change with tone power. We can think of this as now measuring the distance in terms of coupling line widths rather than resonator line widths. This is particularly useful in the presence of a dissipative nonlinearity when the resonator is more strongly $Q_i$-limited than in the data presented here, such as at higher bath temperatures or optical loading.

\section{Nonlinearity Measurements}

\begin{figure}[htbp]%
\centering
\includegraphics[width=0.5\textwidth]{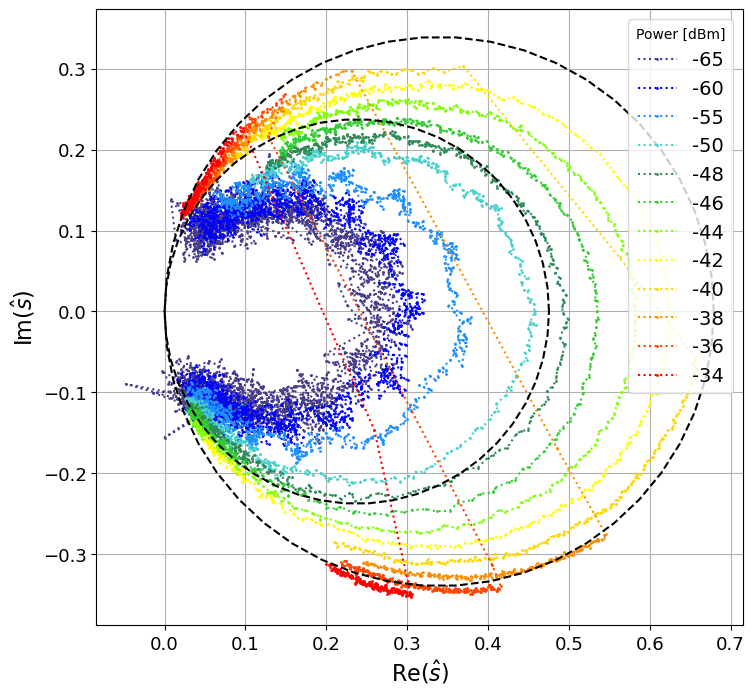}
\caption{Shifted resonance circles as a function of raw tone power for an Al detector. We can immediately see the interplay of the competing nonlinearities. The resonance circle is squashed into an asymmetric "tear-drop" shape at higher powers and the diameter increases significantly with tone power. The dashed circles are included for visual reference.}
\label{fig:Al_IQ}
\end{figure}

\begin{figure}[htbp]
   \begin{center}
\includegraphics[width=\textwidth]{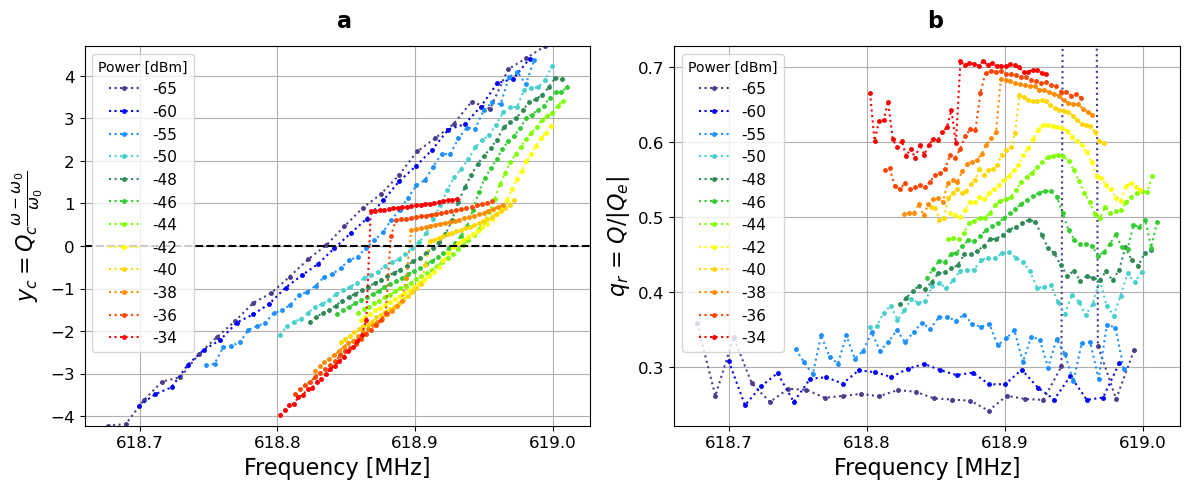}
   \end{center}
\caption{Plots of $y_c$ (a) and $q_r$ (b) from previous Al resonance circles with both reactive and dissipative nonlinearities. You can see a transition between nonlinear effects as the deformation in $y_c$ changes directions with increasing power. Similar plots for TiN are presented in \citenum{duell_superconducting_2024}. }\label{fig:Al_nonlin} 
\end{figure}

Turning our attention to the data, in Figures \ref{fig:Al_IQ} and \ref{fig:Al_nonlin} we can see the behavior of an Al KID under relevant loading conditions of $\sim 5 \text{ pW}$. Even before moving to the nonlinearity parameters, it is possible to see the distortions in the Argand plane that are signatures of both nonlinear kinetic inductance and quasiparticle ``cooling." The resonance circle is expanding and pinching asymmetrically on the sides, while at the highest powers we see nonlinear kinetic inductance causing bifurcation. 

Things become more clear in Figure \ref{fig:Al_nonlin}, which shows $q_r$ and $y_c$. In $y_c$ it is clear that the coupling $Q$ is not changing dramatically as the resonator is being driven between two parallel states. At low powers, we observe a deformation that is similar to that seen from nonlinear kinetic inductance, except that this nonlinearity is occurring in the opposite direction. At higher powers, we observe the nonlinear kinetic inductance begin to flatten the response and eventually drive the resonant frequency back down in the opposite direction. This is corroborated by the plot of $q_r$, which shows the resonator beginning in a strongly $Q_i$-dominated regime and continually driven up into a strongly $Q_c$-dominated regime. We also see which way the resonance is being pushed by the asymmetry in $q_r$, where at lower powers there is a gradual increase in $Q$ as we approach the center frequency, followed by a sharper drop off after passing it, as the resonator snaps back to its low power, low frequency state. At high powers however, the opposite occurs, the resonance is pulled down sharply and then follows the probe tone gradually as is typical of the kinetic inductance non-linearity.  Crucially, we see that the quasiparticle nonlinearity is not quite saturating at higher powers, but is lessened as we become $Q_c$-limited and the nonlinear kinetic inductance drives $y$ in the opposite direction. Lastly, a careful examination of these plots point towards probing the detectors on the high frequency side of resonance as the enhanced slope in the plots of $y_c$ prior to bifurcation reveal how the quasiparticle nonlinearity can serve to amplify the signal from incident power. 

\section{Conclusion}

We showed here a method of using the complex $S_{21}$ data from the full resonator trace to interpret and fit highly nonlinear resonator behavior, though this requires accurate correction of the environmental components (i.e. the cable delay, loop gain, and rotation from impedance mismatches) of the trace. We can use these plots with good constraints on the environment and $Q_c$ values to immediately estimate $f_0$ and $Q$. Looking at Prime-Cam's Al KIDs as described here gives a qualitative picture of how the competing kinetic inductance and quasiparticle nonlinearities affect the line shape under realistic operating conditions from roughly 5--9 pW. At tone powers well below bifurcation due to nonlinear kinetic inductance,  the resonator $Q$ is significantly reduced and $f_0$ is shifted down. With increasing tone power (until near bifurcation), both $Q$ and $f_0$ increase though the resonance circle becomes asymmetrically compressed on the sides off-resonance. By better understanding the behavior in this way, we can develop strategies for how to best account for it in parameter estimation and probe tone placement.

\acknowledgments      
 
The CCAT project, FYST and Prime-Cam instrument have been supported by generous contributions from the Fred M. Young, Jr. Charitable Trust, Cornell University, and the Canada Foundation for Innovation and the Provinces of Ontario, Alberta, and British Columbia. The construction of the FYST telescope was supported by the Gro{\ss}ger{\"a}te-Programm of the German Science Foundation (Deutsche Forschungsgemeinschaft, DFG) under grant INST 216/733-1 FUGG, as well as funding from Universit{\"a}t zu K{\"o}ln, Universit{\"a}t Bonn and the Max Planck Institut f{\"u}r Astrophysik, Garching. ZBH was supported by a NASA Space Technology Graduate Research Opportunity. MN acknowledges support from NSF grant AST-2117631. SW acknowledges support from the Cornell CURES fellowship. 

\bibliography{report} 
\bibliographystyle{spiebib} 

\end{document}